\documentclass[
aps,prd,showpacs,twocolumn,
superscriptaddress]{revtex4-1}

\usepackage{graphicx}
\usepackage{dcolumn}

\usepackage{amsmath}
\usepackage{amssymb}
\usepackage{bm}
\usepackage{color}
\usepackage[normalem]{ulem}
\usepackage[dvipsnames]{xcolor}
\usepackage{hyperref}
\usepackage{tikz}
\hypersetup{
	colorlinks=true, 
	linkcolor=blue,  
	citecolor=cyan,  
}
\newcommand{\be}{\begin{equation}}
\newcommand{\ee}{\end{equation}}
\newcommand{\bea}{\begin{eqnarray}}
\newcommand{\eea}{\end{eqnarray}}

\begin{document}
	
	\title{Particle motion around a static axially symmetric wormhole}
	
	\author{Bakhtiyor Narzilloev}
	\email{nbakhtiyor18@fudan.edu.cn}
	\affiliation{Center for Field Theory and Particle Physics and Department of Physics, Fudan University, 200438 Shanghai, China }
	\affiliation{Akfa University, Kichik Halqa Yuli Street 17, Tashkent 100095, Uzbekistan}
		\affiliation{Ulugh Beg Astronomical Institute, Astronomy St.  33, Tashkent 100052, Uzbekistan}
	
	\author{Daniele Malafarina}
	\email{daniele.malafarina@nu.edu.kz}
	\affiliation{Department of Physics, Nazarbayev University, 53 Kabanbay Batyr avenue, 010000 Astana, Kazakhstan }

\author{Ahmadjon~Abdujabbarov}
\email{ahmadjon@astrin.uz}
\affiliation{Shanghai Astronomical Observatory, 80 Nandan Road, Shanghai 200030, P. R. China}
\affiliation{Ulugh Beg Astronomical Institute, Astronomy St.  33, Tashkent 100052, Uzbekistan}
\affiliation{Institute of Nuclear Physics, Ulugbek 1, Tashkent 100214, Uzbekistan}
\affiliation{National University of Uzbekistan, Tashkent 100174, Uzbekistan}
\affiliation{Tashkent Institute of Irrigation and Agricultural Mechanization Engineers, Kori Niyoziy 39, Tashkent 100000, Uzbekistan}

	\author{Bobomurat Ahmedov}
	\email{ahmedov@astrin.uz}
	
	\affiliation{Ulugh Beg Astronomical Institute, Astronomy St.  33, Tashkent 100052, Uzbekistan}
	\affiliation{National University of Uzbekistan, Tashkent 100174, Uzbekistan}
	\affiliation{Tashkent Institute of Irrigation and Agricultural Mechanization Engineers, Kori Niyoziy 39, Tashkent 100000, Uzbekistan}
	
	\author{Cosimo Bambi}
	\email[Corresponding author: ]{bambi@fudan.edu.cn}
	\affiliation{Center for Field Theory and Particle Physics and Department of Physics, Fudan University, 200438 Shanghai, China }

	\date{\today}

	\begin{abstract}
		
		We consider the properties of a static axially symmetric wormhole described by an exact solution of Einstein's field equations and investigate how we can distinguish such a hypothetical object from a black hole. To this aim, we explore the motion of test particles and photons in the equatorial plane of the wormhole's space-time and compare it with the particle dynamics in the well known space-times of Schwarzschild and Kerr black holes. We show that precise simultaneous measurement of test particle motion and photon motion may provide the means to distinguish the wormhole geometry from that of a black hole.
		
	\end{abstract}

	\maketitle

\section{Introduction}

In recent years, there has been great interest in studying and probing properties of astrophysical black holes due to the direct detection of gravitational waves from close binary mergers by the LIGO and Virgo Collaborations \cite{LIGO16, LIGO16a, LIGO16b, LIGO16c, LIGO16d, LIGO16e, LIGO16f}, observation of the first image of the shadow of a supermassive black hole candidate in the center of elliptic galaxy M87 by the Event Horizon Telescope Collaboration \cite{EHT19a} and the study of the motion of stars and hot spots around the supermassive black hole Sagittarius A* (Sgr A* for short) at the center of the Milky Way galaxy \cite{Ghez98ph,Ghez00ay}.

However, the precision of the measurements collected thus far is still not sufficient to tell whether some other kinds of objects with more exotic properties may also exist in the Universe. Wormholes are unquestionably among the most fascinating theoretical exotic objects that have been studied
\cite{Visser95}. 
Most of the known wormhole space-times, starting with the original work by Einstein and Rosen \cite{Einstein35}, or the well known Ellis solution \cite{Ellis73} and Morris-Thorne solution \cite{Morris88b}, are spherically symmetric (see also \cite{Visser89,Poisson95,Visser03a,Lobo05us}). Interestingly, in recent years the possibility that the supermassive black hole candidates at the center of galaxies may in fact be such wormholes has been considered by several authors \cite{Dai19mse,Li:2014coa,Piotrovich20kae, Zhou16koy, Tripathi19trz, Bambi13jda,Bambi:2021qfo}.
However, it is not widely known that axially symmetric wormholes exist as well \cite{Clement84,Clement83tu,Clement16s}.

Static, axially symmetric vacuum solutions of Einstein's equations play an important role in our understanding of the role played by the mass quadrupole moment in astrophysical compact objects (see \cite{Quevedo90s,Quevedo91s,Pastora94s,Herrera07hz,Bini14tiu,Gurlebeck15xpa,Malafarina20kmk}).
The properties of solutions describing the gravitational field in the exterior of a static massive object with a quadrupole moment have been widely studied
\cite{Curzon25s,Erez59s,Armenti77s,Bonnor92s,Pastora93s,Semerak99s}. 
Among these solutions the most important one is arguably the Zipoy-Voorhees (ZV) metric, also known as the $\gamma$ metric \cite{Zipoy66, Voorhees70}. The importance of the $\gamma$ metric is indicated by the fact that the line element is continuously linked to the Schwarzschild line element through the value of one parameter, $\gamma$, describing the departure from spherical symmetry. For this reason the properties of the $\gamma$ metric, in connection with the possibility of distinguishing the space-time from the Schwarzschild black hole have been extensively studied (see \cite{Stewart82, Herrera05, Papadopoulos81wr,Herrera98rj,Chowdhury11aa,Boshkayev15jaa,Carlos18htf,Abdikamalov19ztb,Toshmatov19qih,Toshmatov19bda}).

In \cite{Gibbons2017} it was shown that known axially symmetric vacuum solutions such as the $\gamma$ metric may be used as the seed to construct new solutions that describe wormholes. These wormhole solutions present ring singularities that can be viewed as the throat of the wormhole. The usual matching procedures can be employed to replace the ring singularities with thin massive rings of negative tension thus allowing for geodesics to cross from one asymptotic region to another.

In this paper we explore the observational properties of the wormhole solution discussed in \cite{Gibbons2017} by using the $\gamma$ metric as the seed and investigate the trajectories of massive particles and photons in the vicinity of the throat. By comparing the results with known results for black holes and the $\gamma$ metric we investigate the possibility of distinguishing such solutions via astronomical observations.

Motion of particles in the vicinity of massive compact objects is a well-known tool to probe the properties of the geometry in the exterior of such compact objects. Particle dynamics in the space-time of various gravitating compact  objects have been extensively studied in the literature; see, for example, our recent papers \cite{Narzilloev:2020b, Narzilloev19, Narzilloev20a, Narzilloev20b, Narzilloev20c, Hakimov17, Narzilloev21, Narzilloev21a}. We also refer the reader to some valuable works devoted to the investigation of the properties of space-time around various wormhole solutions \cite{Abdujabbarov16a, Abdujabbarov09ad}.

The article is organized as follows: In section \ref{wormhole} we describe the axially symmetric wormhole space-time obtained in \cite{Gibbons2017} from the $\gamma$ metric. Sections \ref{particles} and \ref{photons} are devoted to the study of the motion of test particles, photons and gravitational lensing in the given space-time. Finally in section \ref{conc} we briefly discuss the implications of the obtained results for astrophysical compact objects.

Throughout the paper we make use of natural units setting $c=G=1$.

\section{Wormhole metric}\label{wormhole}

The most general line element for a static axially symmetric, vacuum solution of Einstein's equations, in cylindrical coordinates $\{t,\rho,z,\phi\}$ takes the form \cite{Weyl17s,Weyl18,Weyl19s}
\be 
ds^2=-e^{2U}dt^2+e^{2(W-U)}\left(d\rho^2+dz^2\right)+\rho^2e^{-2U}d\phi^2 \; ,
\ee 
and the field equations for the two unknown metric functions $U(\rho,z)$ and $W(\rho,z)$ reduce to
\bea \label{laplace}
&& U{,\rho\rho}+\frac{U_{,\rho}}{\rho}+U_{,zz}=0 \; , \\ \label{quad}
&& W_{,\rho}=\rho(U_{,\rho}^2-U_{,z}^2) \; , \;\; W_{,z}=2\rho U_{,\rho}U_{,z} \; ,
\eea  
where we use the notation $\partial X/\partial x=X_{,x}$. Since $U$ does not depend on $\phi$ the first equation is immediately recognized as a Laplace equation in flat space in cylindrical coordinates. Once a solution of Eq. \eqref{laplace} is given the remaining equations are immediately solved. Therefore, there is a one to one correspondence between solutions of the Laplace equation and static axially symmetric vacuum space-times and, in principle, all metrics of this class are known.

In particular, the $\gamma$ metric is obtained from the solutions of Eqs. \eqref{laplace} and \eqref{quad} given by
\bea \label{gamma-U}
U(\rho,z)&=&\frac{\gamma}{2}\ln\left(\frac{R_++R_--m}{R_++R_-+m}\right)\; , \\ \label{gamma-W}
W(\rho,z)&=&\frac{\gamma^2}{2}\ln\left(\frac{(R_++R_-)^2-4m^2}{4R_+R_-}\right)\; ,
\eea
with 
\be 
R_{\pm}=\sqrt{\rho^2+(z\pm m)^2}\; ,
\ee 
and the Schwarzschild metric is recovered in the limit when $\gamma=1$.
By performing the transformation of coordinates
\bea \label{z}
z&=& r\cos\theta \; ,\\ \label{rho}
\rho&=& \sqrt{r^2-m^2}\sin\theta\; ,
\eea 
we can write the $\gamma$ metric in the form
\bea \label{gamma}
ds^2&=& -\left(\frac{r-m}{r+m}\right)^\gamma dt^2+\left(\frac{r-m}{r+m}\right)^{-\gamma}dl^2 \; ,
\eea 
with
\bea \label{gamma2} \nonumber
dl^2&=& \left(\frac{r^2-m^2\cos^2\theta}{r^2-m^2}\right)^{1-\gamma^2}[dr^2+(r^2-m^2)d\theta^2]+\\
&&+(r^2-m^2)\sin^2\theta d\phi^2 \; .
\eea 
One can show that for the $\gamma$ metric a curvature singularity appears at $r=m$ when $\gamma\neq1$, which corresponds to the infinite red-shift surface. On the other hand the radius $r=m$ corresponds to the event horizon of the Schwarzschild black hole (BH) when $\gamma=1$.
Notice that the radial coordinate $r$ employed here is not the Schwarzschild-like coordinate $r_s$ that is usually employed in the description of the $\gamma$ metric. This can be easily seen by taking $\gamma=1$, which must reduce to the Schwarzschild geometry. From this we see that $r$ is simply a translation of the Schwarzschild-like coordinate given by 
\be\label{coord}
r=r_s-m\; ,
\ee 
for which the singularity of the $\gamma$ metric is shifted from $r_s=2m$ to $r=m$.

The procedure to obtain the wormhole space-time starting with the $\gamma$ metric was already presented in \cite{Zipoy66, Voorhees70} and it involves a rotation of the parameters $m$ and $\gamma$ in the complex plane. By applying the transformation
\bea 
m&=&i\mu \; ,\\
\gamma&=&i\sigma \; ,
\eea 
the field equations become complex but, remarkably, the solutions remain real. In fact we obtain
\bea
\label{e2.4.2}
U&=&\sigma \tan^{-1}\left(\frac{r}{\mu}\right) \; , \\
W&=&\frac{\sigma^2}{2}\ln\left(\frac{r^2+\mu^2\cos^2\theta}{r^2+\mu^2}\right)\; ,
\eea
and the space-time metric that describes wormholes is then given by the line element
\begin{eqnarray}
\label{metric}
ds^2=-e^{2U}dt^2+e^{-2U}\Bigg[\left(\frac{r^2+\mu^2\cos^2\theta}{r^2+\mu^2}\right)^{1+\sigma^2} \\\nonumber
\times[dr^2+(r^2+\mu^2)d\theta^2]+(r^2+\mu^2)\sin^2\theta d\phi^2\Bigg] \; .
\end{eqnarray}

This is the standard form of the oblate Zipoy-Voorhees solutions that were already given in \cite{Zipoy66,Voorhees70}. The metric in this form describes a wormhole with two asymptotically flat regions for $r\rightarrow\pm\infty$ which are connected by a throat at $r=0$. Notice that the metric is Ricci flat and the Kretschmann scalar $\mathcal{K}=R_{\alpha\beta\gamma\epsilon}R^{\alpha\beta\gamma\epsilon}$ is given by 
\begin{equation}
\label{e2.4.3}
\mathcal{K}=\frac{64 \mu ^2 \sigma ^2 e^{4 \sigma  \tan^{-1} \left(r/\mu \right)} \left(\frac{\mu ^2 \cos ^2\theta +r^2}{\mu ^2+r^2}\right)^{-2 \sigma ^2}\mathcal{A}}{\left(\mu ^2+r^2\right)^2 \left(\mu ^2 \cos 2 \theta +\mu ^2+2 r^2\right)^3} \; ,
\end{equation}
where
\begin{eqnarray}
\label{e2.4.4}
\begin{aligned}
\mathcal{A}&=\mu ^4 \left(\sigma ^4+5 \sigma ^2+1\right)+6 r^4-12 \mu  \sigma  r^3
\\\nonumber
&+3 \mu ^2 \left(3 \sigma ^2+2\right) r^2-3 \mu ^3 \sigma  \left(\sigma ^2+3\right) r -\mu ^2 \cos 2 \theta \\
&\times \left[\mu ^2 \left(\sigma ^4-\sigma ^2+1\right)+3 \sigma ^2 r^2
-3 \mu  \sigma  \left(\sigma ^2-1\right) r\right] \; .
\end{aligned}
\end{eqnarray}
One can easily check from the form of the Kretschmann scalar to see that on the equatorial plane ($\theta=\pi/2$) the denominator of the expression tends to zero when $r\rightarrow0$. Therefore, the space-time possesses a ring singularity at $r=0$ in the equatorial plane. By a standard cut and paste procedure, the ring singularity can be replaced by a massive thin ring with negative tension that joins the two wormhole regions \cite{Gibbons2017}.  Also it is easy to notice that the metric in the form given in equation \eqref{metric} does not reduce to Minkowski space-time for $r\rightarrow\pm\infty$; however, this may be achieved by making a simple rescaling of the coordinates $t$ and $r$ and the geometry is asymptotically flat.

One should point out that the line element \eqref{gamma} can also be made a wormhole using a cut and paste procedure with another identical line element at the singularity, which must then be replaced by a matter distribution. However, this is not as natural a construction as the one obtained in \cite{Gibbons2017}, where the complex rotation of the parameters gives rise to quadratic equations for the functions involved. Then by taking the positive and negative roots one sees that these can be glued continuously at the wormhole's throat, thus effectively allowing for the radial coordinate to run from $-\infty$ to $+\infty$. This makes clear that the metric functions obtained by the complex rotation describe two separate asymptotically flat regions connected by a throat, i.e., a wormhole, where both portions of the geometry are obtained from the same symmetry transformation.

From the above one can see that the structure of the wormhole space-time with line element \eqref{metric} differs from that of the $\gamma$ metric given by the expressions \eqref{gamma} and \eqref{gamma2}. This can be made more clear by investigating radial photon motion in the two space-times. 
We know that for the $\gamma$ metric the singular surface $r=m$ is infinitely redshifted when $\gamma>1$ while the singularity is naked, i.e. photons take a finite amount of time to reach any observer, when $\gamma<1$. On the other hand, a similar investigation for the ZV wormhole shows that the curvature throat at $r=m$ is naked for any value of $\sigma$.

\subsection{Komar integrals}
How would distant observers measure the mass of the ZV wormhole? In order to answer this question we employ the Komar integrals, which are a standard tool to determine the mass and the angular momentum as seen by a faraway observer in a given geometry. These integrals are defined for stationary and axially symmetric space-times, which must be asymptotically flat. Then mass and angular momentum of the gravitating massive object for faraway observers are given by 
\begin{equation}
\label{2.5.1}
\begin{aligned}
M&=&\frac{1}{4\pi }\int_{\partial\Sigma} dr^2\sqrt{{\rm g}^{(2)}}n_\mu\chi_\nu\nabla^\mu K^\nu \; ,\\
J&=&-\frac{1}{8\pi }\int_{\partial \Sigma}dr^2\sqrt{{\rm g}^{(2)}}n_\mu\chi_\nu\nabla^\mu R^\nu \; ,
\end{aligned}
\end{equation}
where $n_\mu$ is a time-like normal unit vector ($n_\mu n^\mu=-1$), while $\chi_\nu$ is a space-like normal vector ($\chi_\mu \chi^\mu=1$). Also, $K^\nu$ is the timelike killing vector associated with time invariance and $R^\nu$ is the Killing vector associated with rotation invariance about the symmetry axis, and $\sqrt{{\rm g}^{(2)}}$ is the determinant of the metric on the two-dimensional surface of constant $t$ and constant $r$ at infinity. Since the space-time is static, it is clear that one must have $J=0$.
To evaluate $M$ we use the normal vectors $n_\mu$ and $\chi_\nu$ which are given by 
\bea
\label{2.5.2}
n_\mu&=&\left(-e^{-U},0,0,0\right) \; , \\
\chi_\mu&=&\left(0,e^{-U}\left[\frac{r^2+\mu^2\cos^2\theta}{r^2+\mu^2}\right]^{(1+\sigma^2)/2},0,0\right) \; ,
\eea	 
so we get
\begin{equation}
\label{2.5.8}
n_{\mu}\chi_\nu\nabla^\mu K^\nu=\frac{\mu  \sigma  e^{2 \sigma  \tan ^{-1}\left(r/\mu \right)} }{\mu ^2+r^2}\left(\frac{\mu ^2 \cos ^2\theta +r^2}{\mu ^2+r^2}\right)^{-\frac{\sigma ^2+1}{2}} \; .
\end{equation}

The line element for the surface of constant $t$ and $r$ is
\begin{eqnarray}
\label{2.5.9}
{\rm g}_{\alpha\beta}&=&\frac{\mu ^2+r^2}{e^{2 \sigma  \tan ^{-1}\left(r/\mu \right)}} \left(\frac{\mu ^2 \cos ^2\theta +r^2}{\mu ^2+r^2}\right)^{\sigma ^2+1} d\theta^2+
\\\nonumber
&+&\sin ^2\theta  \left(\mu ^2+r^2\right) e^{-2 \sigma  \tan ^{-1}\left(r/\mu \right)}d\phi^2 ,
\end{eqnarray}
from which we obtain ${\rm g}=\det({\rm g}_{\alpha\beta})$ as
\begin{eqnarray}
\label{2.5.10}
{\rm g}=\frac{\sin \theta  \left[\mu ^2 \cos ^2\theta +r^2\right]}{e^{2 \sigma  \tan ^{-1}\left(r/\mu \right)}}  \left(\frac{\mu ^2 \cos ^2\theta +r^2}{\mu ^2+r^2}\right)^{\frac{\sigma ^2-1}{2}} \; .
\end{eqnarray}
Therefore the Komar integral \eqref{2.5.1} reduces to
\begin{eqnarray}
\label{2.5.11}
M&=&\frac{\mu\sigma}{4\pi }\int^{2\pi}_0\int^{\pi}_0\sin\theta d\theta d\phi=\mu\sigma.
\end{eqnarray}
This is consistent with the interpretation of the active gravitational mass of the $\gamma$ metric \eqref{gamma}, which is given by $M=m\gamma$.

\section{Test particle motion}\label{particles}

In this section we aim to investigate whether and how a distant observer that detects a compact object of mass $M$ would be able to tell if the object is a black hole (Kerr or Schwarzschild), a deformed exotic compact object (such as the $\gamma$ metric) or a wormhole (in this case the axially symmetric one given by the ZV solution).
Toward this aim we investigate the motion of test particles in the space-time of the wormhole with the metric given in equation (\ref{metric}) and compare it to the other sources. 

To begin with, we consider the comparison with the Schwarzschild black hole since both are static solutions and show that the two space-times produce different orbits for particles. We then compare the obtained results to that in the Kerr space-time since both space-times have one parameter in addition to the mass and investigate the conditions for these two metrics to mimic each other in terms of particle dynamics. Notice that in making the comparisons we will need to employ the coordinates used in the line element \eqref{metric} instead of the usual Boyer-Lindquist coordinates. 

For the derivation of the equations of motion for test particles we use the well known Hamilton-Jacobi equation that reads
\be
g^{\alpha\beta}\frac{\partial {\cal S}}{\partial
	x^{\alpha}}\frac{\partial {\cal S}}{\partial
	x^{\beta}} = -k\, ,
\label{mHJ}
\ee
where $S$ defines the action for the test particle, $x^\alpha$ are the coordinates and $k=m_p^2$, i.e. the square of the mass of the test particle. Notice that the equation also holds for massless particles, i.e. photons, for which $k=0$. Since the metric of the given space-time is independent from the coordinates $t$ and $\phi$, the particle orbiting the wormhole has two conserved quantities related to time translations and rotations, namely, the energy $E$ and angular momentum $L$. As a consequence the action for the particle can be written in the following form
\begin{eqnarray}
S=-E t + L \phi + S_\theta + S_r\ .
\end{eqnarray}
Here $S_\theta$ and $S_r$ are functions of $r$ and $\theta$ only.

The equation of motion for the test particle then reads
\begin{widetext}
\be \label{eom}
-\kappa=\frac{  e^{2 U} \csc ^2\theta}{\mu^2+r^2}\mathcal{L}^2 +\frac{e^{2 U}}{\mu ^2+r^2}
	\left(\frac{\partial S_\theta}{d\theta}\right)^2 \left(\frac{\mu ^2 \cos ^2\theta +r^2}{\mu ^2+r^2}\right)^{-\sigma
		^2-1}+e^{2 U} \left(\frac{\partial S_r}{dr}\right)^2 \left(\frac{\mu ^2 \cos ^2\theta +r^2}{\mu
	^2+r^2}\right)^{-\sigma ^2-1}- e^{-2 U}\mathcal{E}^2,
\ee
\end{widetext}
where we used the notation $\mathcal{E}=E/m_p$ and $\mathcal{L}=L/m_p$ for the energy and angular momentum of the particle with unit mass, $\kappa=1$ for massive particles, and $\kappa=0$ for massless particles.

\begin{figure}
 	\centering\includegraphics[width=0.958\linewidth]{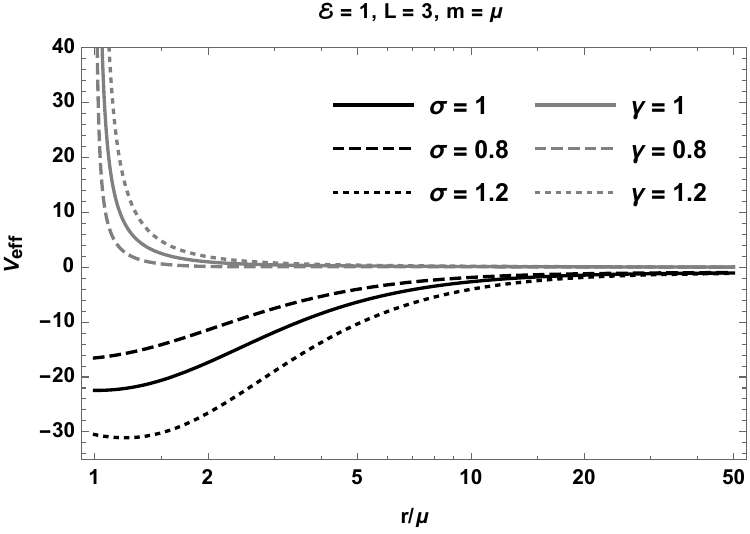}
	\caption{The radial dependence of the effective potential $V_{eff}(r)$  for massive test particles in the equatorial plane of the ZV wormhole, with given energy $\mathcal{E}$ and angular momentum $\mathcal{L}$ per unit mass, for a fixed $\mu$ and various values of the deviation parameter $\sigma$ (black lines) is compared to the effective potential for massive test particles in the equatorial plane of the $\gamma$ metric for a fixed $m$ and various values of the deformation parameter $\gamma$ (gray lines). Notice that the case $\gamma=1$ corresponds to the Schwarzschild black hole.  \label{Veff}}
 \end{figure}

In the following we will focus on the motion of test particles on the equatorial plane $\theta=\pi/2$ thus setting $\dot{\theta}=0$. Then the equation of motion can be written in the form $\dot{r}^2+V_{eff}(r)=0$ where the effective potential $V_{eff}$ is given by

\begin{eqnarray} \label{Veff2}
V_{eff}(r)=1-\frac{\mathcal{E}^2}{ e^{2 \sigma  \tan ^{-1}\left(r/\mu\right)}}+\mathcal{L}^2\frac{ e^{2 \sigma 
		\tan ^{-1}\left(r/\mu \right)}}{\mu ^2+r^2}\; .
\end{eqnarray}
The radial behavior of the effective potential is shown in Fig. ~\ref{Veff} for different values of the deviation parameter $\sigma$ in comparison with the corresponding cases for the $\gamma$ metric and Schwarzschild ($\gamma=1$). 
As expected, at large distances the behaviors tend to become increasingly similar. 
At the same time, the wormhole geometry is significantly different from the black hole and the $\gamma$ metric at shorter distances for every value of the parameter $\sigma$. In particular, and unlike the case of the $\gamma$ metric, the case of $\sigma=1$ does not reduce to a known spherically symmetric geometry.
This suggests that from the the motion of test particles around a central gravitating object it is, in principle, possible to distinguish the wormhole from a black hole or a static and axially symmetric compact object.

We aim now at studying circular orbits, as they are well suited to approximating the orbits of particles in accretion disks, and, in particular, we wish to determine the value of the innermost radius for which stable circular orbits are allowed. For the trajectory of the particle to be circular the conditions $\dot{r}=\ddot{r}=0$ must be satisfied. These conditions translate into the corresponding conditions for the effective potential, namely
 \begin{eqnarray}\label{c1}
 V_{eff}(r)=V_{eff}'(r)=0\, ,
 \end{eqnarray}
from which one can obtain expressions for the energy and angular momentum of the test particle on circular orbit as

 \bea 
 \mathcal{E}&=& \sqrt{\frac{r-\sigma\mu}{r-2\sigma\mu}}e^{\sigma\tan^{-1}(r/\mu)}\; ,\\
 \mathcal{L}&=&\sqrt{\frac{\sigma\mu(r^2+\mu^2)}{r-2\sigma\mu}}e^{-\sigma\tan^{-1}(r/\mu)}\; .
 \eea 
 
\begin{figure*}
    \centering\includegraphics[width=0.49\linewidth]{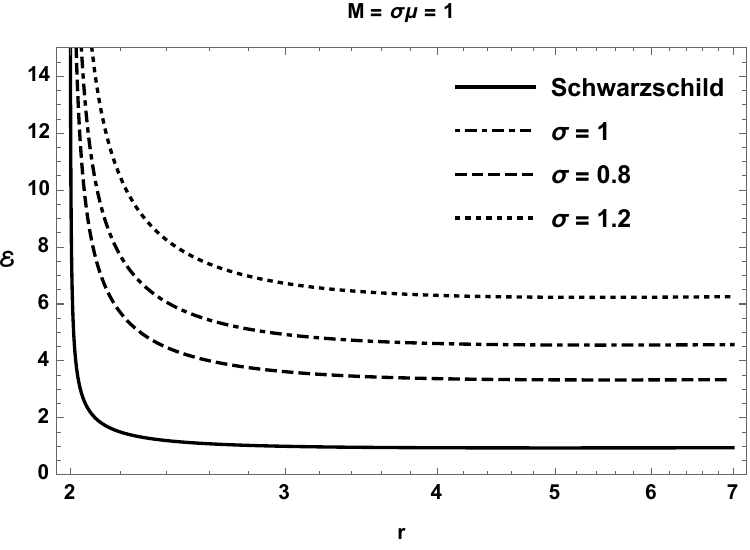}
	\centering\includegraphics[width=0.49\linewidth]{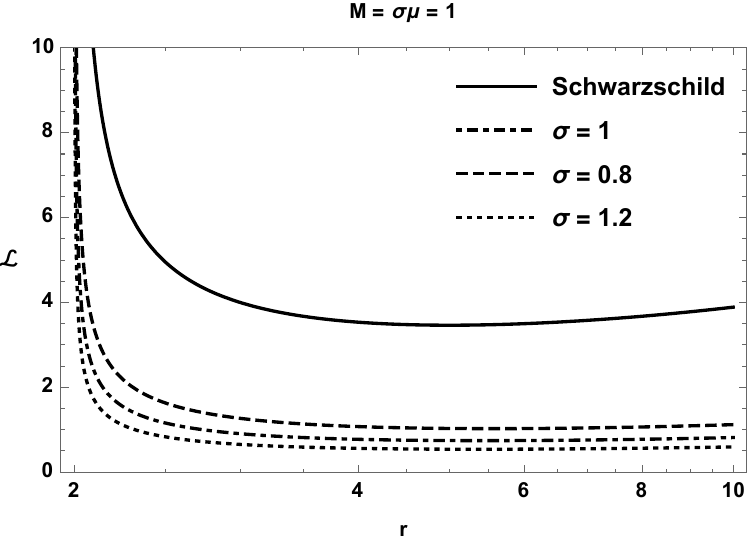}
	\centering\includegraphics[width=0.49\linewidth]{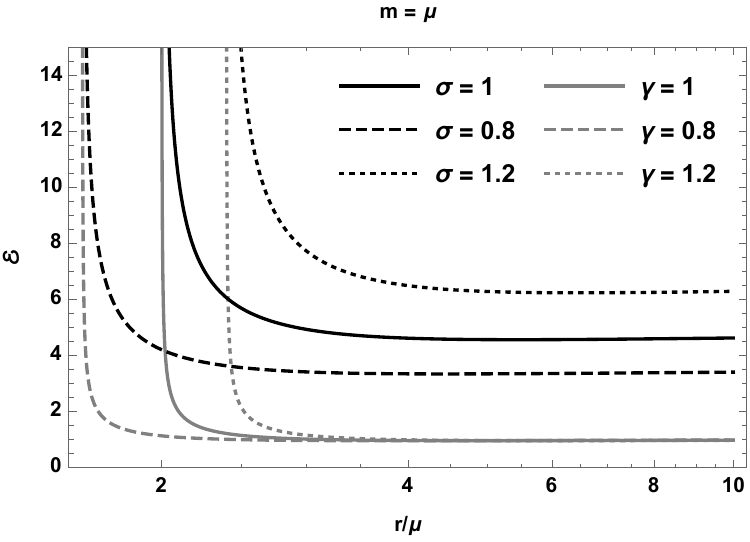}
	\centering\includegraphics[width=0.49\linewidth]{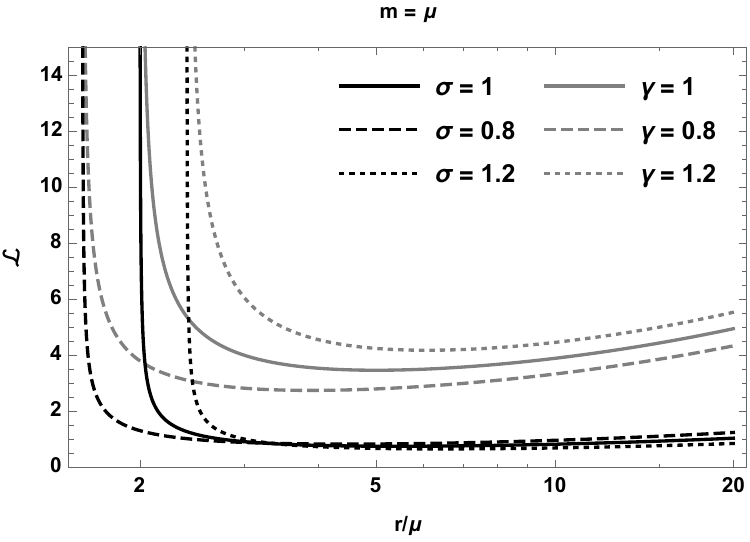}
	\caption{Top row: radial dependence of the energy (left panel) and angular momentum (right panel) for test particles moving on circular orbits for a fixed gravitational mass $M=\sigma\mu=1$ in the ZV wormhole geometry with various values of $\sigma$ compared to the Schwarzschild space-time (solid line). 
	Bottom row: radial dependence of the energy (left panel) and angular momentum (right panel) for test particles moving on circular orbits in the 
	ZV wormhole geometry compared to the $\gamma$ metric for fixed values of $\mu=m$ and various values of $\sigma$ and $\gamma$ (with $\gamma=1$ corresponding to Schwarzschild). Remember that the usual Schwarzschild radial coordinate is given by $r_s=r+M$, therefore the photon sphere for Schwarzschild is given by $r=2M$. We notice that the energy of particles on circular orbits is always larger for the wormhole with respect to the black hole. Correspondingly the angular momentum is always lower for the wormhole with respect to the black hole. \label{E}}
\end{figure*}

 The radial behaviors of the energy and angular momentum of the particles are plotted in Fig.~\ref{E}. It is clearly seen that particles on circular orbits have larger energy with respect to corresponding particles in the Schwarzschild space-time and that increasing the deviation parameter $\sigma$ results in an increase of the energy of the particle at any given circular orbit. Similarly, test particles or circular orbits in the equatorial plane of the ZV wormhole have lower angular momentum with respect to particles orbiting the Schwarzschild black hole and an increase of the value of $\sigma$ results in a reduction of the angular momentum of the particle. 
 Also, notice that, as happens for Schwarzschild, the energy and angular momentum per unit mass of the test particles on circular orbits diverge in the limit $r\rightarrow 2\sigma\mu$ which corresponds to the photon capture radius (remember that for Schwarzschild the photon capture orbit is given by $r_s=3M$ which corresponds to $r=2M$).

The minimum of the effective potential, for given values of $\mathcal{E}$ and $\mathcal{L}$, corresponds to the radius of the stable circular orbit. Then it is easy to see that circular orbits can exist only within certain ranges of $r$ and the smallest radius, called the innermost stable circular orbit (ISCO), is important in the study of astrophysical sources because it determines the distance from the central objects where the accretion disks end. In the case of the ZV wormhole the location of the ISCO radius depends on the value of the parameter $\sigma$.
Such behavior is similar to the case of the Kerr metric where the location of the ISCO depends on the angular momentum parameter $a$; in the case of Kerr an increase of the rotation parameter $a$ for a co-rotating disk makes the ISCO radius move closer to the black hole. 
Using the equations \eqref{c1} to obtain $\mathcal{E}$ and $\mathcal{L}$ for a particle in a circular orbit, the condition in which the particle is at the ISCO is then given by
\begin{eqnarray}\label{ISCO_eqn}\nonumber
V_{eff}''(r)&=&0\ ,
\end{eqnarray}
which gives the quadratic equation 
\be 
r^2-6\sigma\mu r+(4\sigma^2-1)\mu^2=0 \; ,
\ee 
from which we find the ISCO radius as
\be \label{sigma-ISCO}
r_{ISCO}=3\sigma\mu\pm\sqrt{5\sigma^2\mu^2+\mu^2} \; .
\ee
It is useful to point out here that when we make an inverse mapping i.e. $\sigma \mu \rightarrow\gamma m$  
we obtain the expression for the ISCO radius for the $\gamma$ metric that, in our coordinates, reads $r_{ISCO} = 3 \gamma m \pm \sqrt{5 \gamma^2 m^2 - m^2}$. As shown, for example, in \cite{Toshmatov19qih} the $\gamma$ metric has two locations for the marginally stable circular orbits, which we may call the inner and outer ISCO. However, for certain values of $\gamma$ one of these radii is not physical. For example, when $\gamma=1$ we have $r_{ISCO}=5m$ and $r_{ISCO}=m$, so that the first solution gives the ISCO radius of the Schwarzschild BH while the second one gives the Schwarzschild horizon. 
In general for oblate sources ($\gamma>1$) only the outer ISCO is physical as the inner ISCO is located below the singularity. On the other hand, for prolate sources the inner ISCO can become physical. In fact for $1/\sqrt{5}<\gamma<1/2$ there are two radii at which particles can ciruolarize, which suggests the occurrence of repulsive effects in the vicinity of the singularity (see \cite{Toshmatov19qih} for details).

As in the case of the Kerr space-time, we see that the location of the ISCO of the ZV wormhole depends on $\sigma$. For the ZV wormhole metric (\ref{metric}) a decrease in the value of the deviation parameter $\sigma$ produces an effect similar to the increase of the rotation parameter $a$ of a Kerr black hole, i.e. it makes the ISCO radius smaller.

\begin{figure*}[ttt]
    \centering\includegraphics[width=0.49\linewidth]{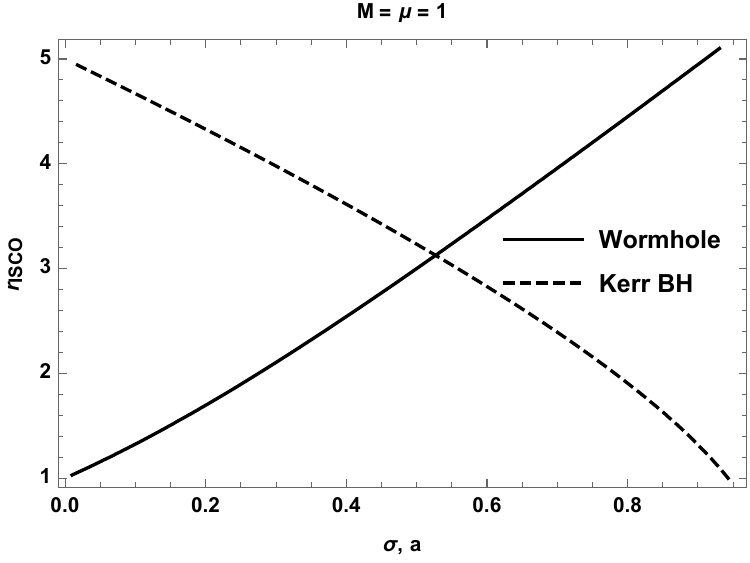}
	\centering\includegraphics[width=0.5\linewidth]{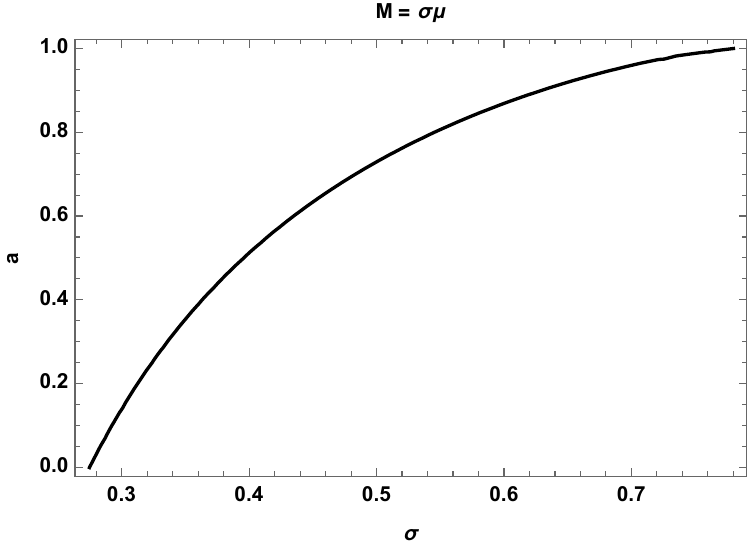}
\caption{Comparison between the ISCO in the ZV wormhole space-time and Kerr black hole. Left panel: value of the ISCO radius (in the rescaled coordinates) as a function of $\sigma$ for the ZV wormhole (solid line) and $a$ for Kerr (dashed line).
Right panel: degeneracy between the ZV wormhole and the Kerr black hole. The curve shows how the deviation parameter of the ZV wormhole can mimic the effect of the spin of the Kerr black hole providing the same ISCO location of test particles which in turn defines the inner edge of the accretion disk around the astrophysical compact object. Note that the plot is obtained for sources with the same gravitational mass $M$ as seen by distant observers.
\label{ISCO}}
\end{figure*}

With the use of the three conditions above one may obtain the dependence of the ISCO from the deviation parameter as in equation \eqref{sigma-ISCO}. This is shown in the left panel of Fig.~\ref{ISCO} in comparison to the case of the Kerr black hole, where the ISCO depends on the angular momentum parameter $a$. 
We notice that the ranges of radii for the ZV wormhole for small values of $\sigma$ are comparable to the radii allowed for the ISCO in the Kerr geometry with co-rotating disk.  
Knowing the dependence of the ISCO radius on the rotation parameter $a$ of the Kerr metric one can plot the degeneracy between the rotation parameter $a$ and the deviation parameter $\sigma$ of the wormhole such that the values of the ISCO radii coincide.

{However, one needs to take into account the fact that in GR the choice of coordinates is arbitrary and has no real physical meaning. Therefore, it makes sense to compare the results for values of scalar quantities related to the characteristic orbits. For example, in our case one can choose the distance of the ISCO orbit from the origin, which is defined as the radius of a circle with the circumference given by 
$$
l_\phi=\int_0^{2 \pi} ds_\phi \ ,
$$ 
where $ds_\phi=\sqrt{g_{\phi \phi}}|_{r=r_{ISCO}} d\phi$ is the line element for constant $r$, $\theta=\pi/2$, and $t$. Since axially symmetric space-times do not depend on the angular coordinate $\phi$ the ISCO location becomes
\begin{eqnarray}\label{R}
R_{ISCO}=\frac{l_\phi}{2 \pi}=\sqrt{g_{\phi \phi}}|_{r=r_{ISCO}}.
\end{eqnarray}
With this definition one can compare the properties of the ZV wormhole and Kerr black hole  in a manner that does not depend on the coordinate choice. The relation between the location of the ISCOs in the Kerr geometry and ZV wormhole is shown in the right panel of Fig.~\ref{ISCO}. It is immediately clear that the effect of the deviation parameter of the ZV wormhole can mimic the spin of the Kerr black hole producing the same radius for the inner edge of the accretion disk around an astrophysical compact massive object. }

Therefore, we conclude from the observation of only the accretion disk around a compact object of known mass $M$, that if we are able to determine the value of the ISCO radius, we may not be able to determine with certainty that the object must be a Kerr black hole, since there is a value of $\sigma$ for which the ZV wormhole would exhibit the same radius of the ISCO.

\begin{figure}[hhh]
	\centering\includegraphics[width=0.95\linewidth]{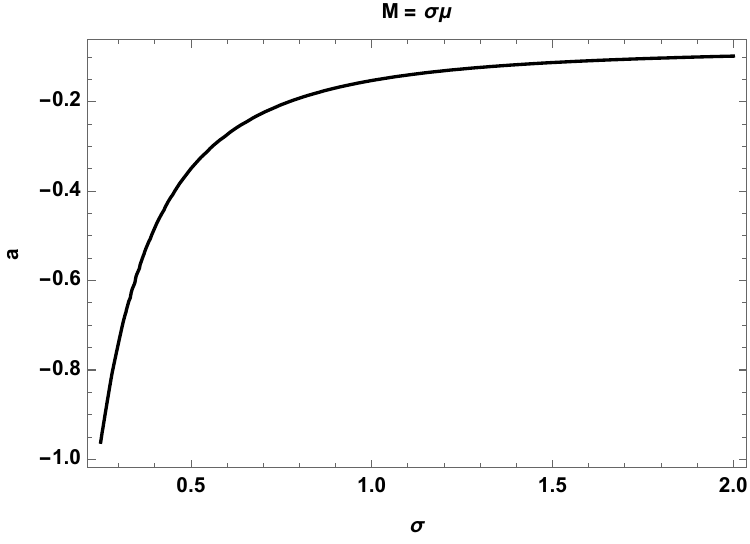}
\caption{Degeneracy plot between the spin of the Kerr BH and the deviation parameter of the ZV wormhole providing the same radiative efficiency.\label{REa_d}}
\end{figure}

{ To complicate things further, observations cannot measure the ISCO location directly but instead measure other properties related to the intensity or spectrum of the light emitted by the disk. For example, observations can determine the radiative efficiency of the system at the ISCO~\cite{Kong:2014wha}. Therefore it is worth checking to see whether the ZV wormhole can provide the same radiative efficiency as the Kerr black hole. The radiative efficiency of the disk around a massive object is given by the expression 
$$
\eta=1-\mathcal{E}_{ISCO}\; ,
$$ 
where $\mathcal{E}_{ISCO}$ is the specific energy of a test particle determined at the ISCO. One can then follow a similar procedure as before to determine the energy at the ISCO, which is clearly independent of the coordinate system. 
In Fig.~\ref{REa_d} we show the values of the Kerr angular momentum that produce the same radiative efficiency for the ISCO as the ZV wormhole with a given value of $\sigma$. 
Notice that the result does not reproduce the one obtained for the matching values of the ISCO location in the right panel of Fig.~\ref{ISCO}. This suggests that a given value of $\sigma$ may produce the same location for the ISCO but different radiative efficiency for the disk. Therefore, this feature may allow one to distinguish the ZV wormhole from a Kerr black hole if one is able to make independent measurements of the location of the inner edge of the accretion disk and the radiative efficiency of the same source. In Fig.~\ref{REa_d} we can also see that the radiative efficiency of a disk surrounding the Schwarzschild BH, i.e. $a=0$, or that of a slowly spinning Kerr black hole may not be reproduced by the ZV wormhole for any value of $\sigma$.}

\begin{figure*}[ttt]
	\centering\includegraphics[width=0.48\linewidth]{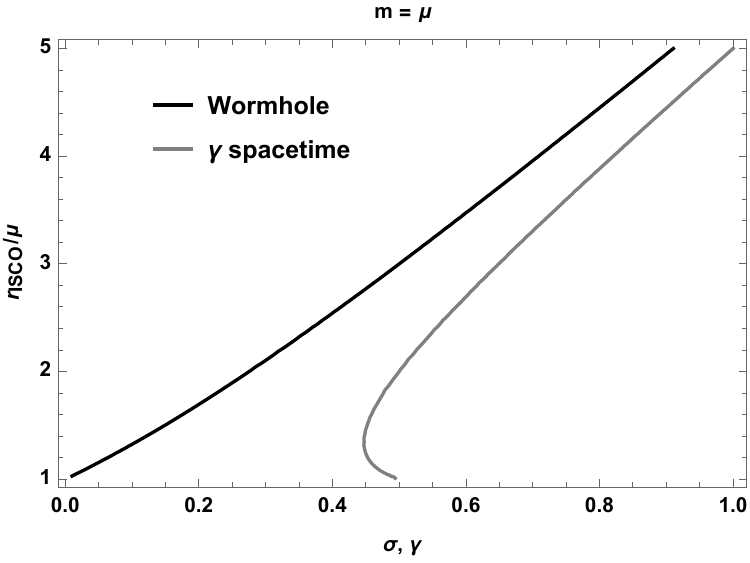}
	\centering\includegraphics[width=0.49\linewidth]{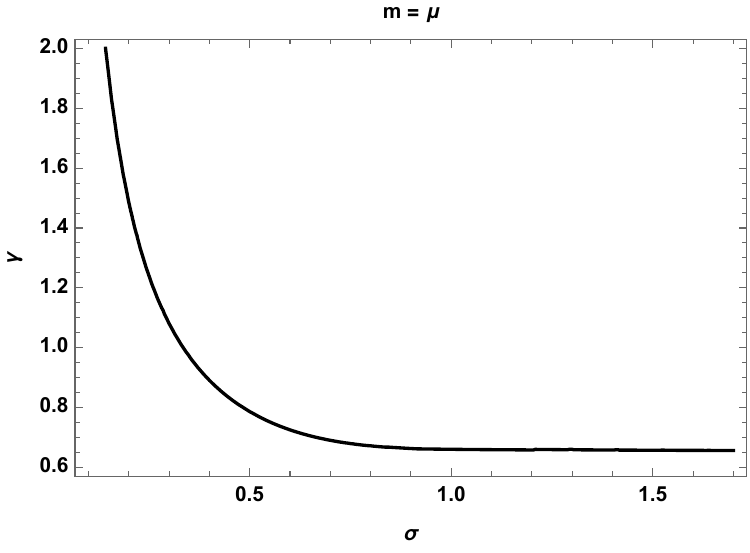}
\caption{Comparison between the ISCO in the ZV wormhole space-time and the $\gamma$ metric. Left panel: value of the ISCO radius (in the rescaled coordinates) as a function of $\sigma$ for the ZV wormhole and $\gamma$ for the $\gamma$ metric. Right panel: degeneracy between the parameters $\sigma$ and $\gamma$, i.e. the values of the two parameters that produce the same value for the ISCO.\label{ISCO2}}
\end{figure*}

For completeness we also compare the ZV wormhole with the static and axially symmetric geometry of the $\gamma$ metric. In Fig.~\ref{ISCO2} we compare the ISCO radii around the ZV wormhole and the $\gamma$ metric. It is interesting to notice that, unlike in the case of the $\gamma$ metric, the ISCO radius exists for all values of $\sigma>0$ and is always larger than the corresponding ISCO radius for the $\gamma$ metric. Therefore, the curious situation of a range of values of $\gamma$ with two separate regions of allowed stable circular orbits does not occur in the wormhole geometry.

{ This degeneracy may also be illustrated from the definition of the distance of the ISCO from the center using equation \eqref{R}.
In the right panel of Fig.~\ref{ISCO2} we can see that the deviation parameter of the ZV wormhole can determine an ISCO that mimics the corresponding one in the $\gamma$ metric for certain values of $\gamma$. However, the scenarios given by certain values of $\gamma<1$, which produce the peculiar cases of no ISCO radius or two regions of stable circular orbits (see \cite{Toshmatov19qih}) cannot be reproduced in the wormhole space-time.}

\begin{figure}[h]
	\centering\includegraphics[width=0.95\linewidth]{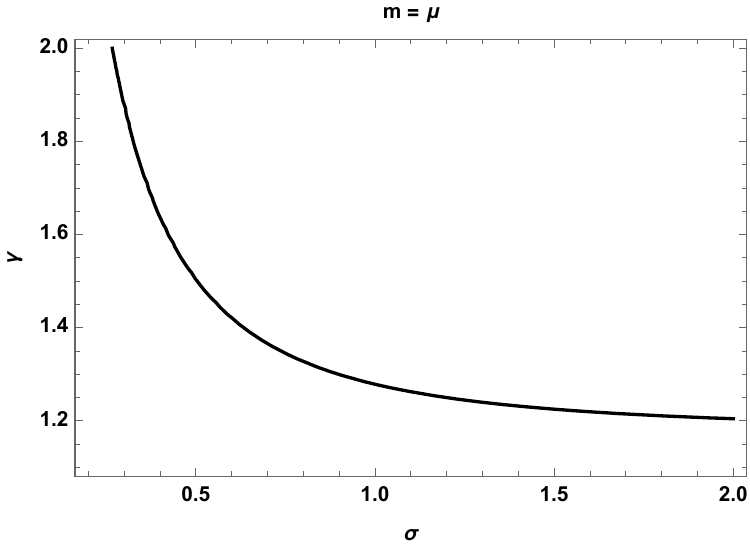}
\caption{Relation between the deviation parameters of ZV wormhole and the $\gamma$ metric providing the same radiative efficiency from the accretion disk surrounding the massive object. Notice that, as happens in the comparison with the Kerr BH in Fig.~\ref{REa_d}, the radiative efficiency of a disk surrounding the Schwarzschild BH, i.e. $\gamma=1$, may not be reproduced for any value of $\sigma$. \label{REd_d}}
\end{figure}

{As before, we can check in what range the deviation parameter of the ZV wormhole can mimic that of the $\gamma$ metric for the radiative efficiency of these two sources. In Fig.~\ref{REd_d} we can see that the degeneracy of the radiative efficiency produces a behavior similar to that displayed in the case of the ISCO radius. However, it is clear that the range of values of $\gamma$ for which the two geometries produce comparable radiative efficiencies is different, as it starts at $\gamma\simeq1.2$ in this case while in the ISCO location case it starts at $\gamma\simeq0.6$. As in the case of the Kerr BH, we can conclude that the case in which $\gamma=1$, or, equivalently, the Schwarzschild case, may not be mimicked by the ZV wormhole via observations of the radiative efficiency of the accretion disk.}

\section{Photon motion and lensing}\label{photons}

We shall now deal with the motion of photons in the wormhole space-time given by equation (\ref{metric}). Again, we use the Hamilton-Jacobi equation \eqref{mHJ}, this time for massless particles, thus setting $k=0$. 
The effective potential for photons moving on the equatorial plane then becomes
\be \label{Veff3}
V_{eff}=-\frac{\mathcal{E}^2}{ e^{2 \sigma  \tan ^{-1}\left(r/\mu\right)}}+\frac{\mathcal{L}^2 e^{2 \sigma 
		\tan ^{-1}\left(r/\mu \right)}}{\mu ^2+r^2}\; .
\ee

\begin{figure}[h]
	\centering\includegraphics[width=0.958\linewidth]{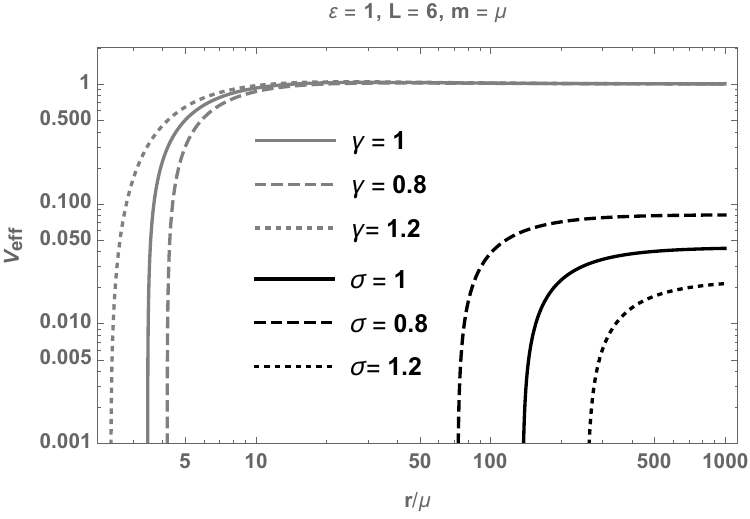}
	\caption{The radial dependence of the effective potential $V_{eff}(r)$  for massless particles in the equatorial plane of the ZV wormhole, with given energy $\mathcal{E}$ and angular momentum $\mathcal{L}$, for fixed $\mu$ and various values of the deviation parameter $\sigma$ (black lines) is compared with the corresponding effective potential for the $\gamma$ metric for a fixed $m$ and various values of the deformation parameter $\gamma$ (gray lines). Notice that the case $\gamma=1$ corresponds to the Schwarzschild black hole.
	\label{Veffp}}
\end{figure}

The radial behavior of $V_{eff}(r)$ is shown in Fig.~\ref{Veffp} relative to the Schwarzschild black hole. It is immediately seen that the two geometries produce very different effective potentials regardless of the value of $\sigma$. 

Our interest in the motion of photons lies in determining the photon capture radius $r_{ph}$ at which massless particles on the equatorial plane are circularized. 
From the geodesic equation $\ddot{x}^{\mu}+\Gamma^{\mu}_{\alpha\beta} \dot{x}^{\alpha}\dot{x}^{\beta}=0$ this is obtained by imposing the conditions $\ddot{r}=\dot{r}=\dot{\theta}=0$. In turn this reflects on the conditions on the effective potential $V_{eff}=V_{eff}'=0$. After a straightforward calculations we obtain the photon capture radius simply as
\be
r_{ph}=2\sigma\mu \; .
\ee
Remembering the relation \eqref{coord} between the radial coordinate used for the ZV wormhole and the Schwarzschild-like coordinate $r_s$ and the gravitational mass of the $\gamma$ metric, $M=m\gamma$. we see that the photon capture radius is located at the same position, i.e. $r_{ph}=2M$, where $M=\sigma\mu$ for the ZV wormhole and $M=m\gamma$ for the $\gamma$ metric.

\begin{figure*}
    \centering\includegraphics[width=0.49\linewidth]{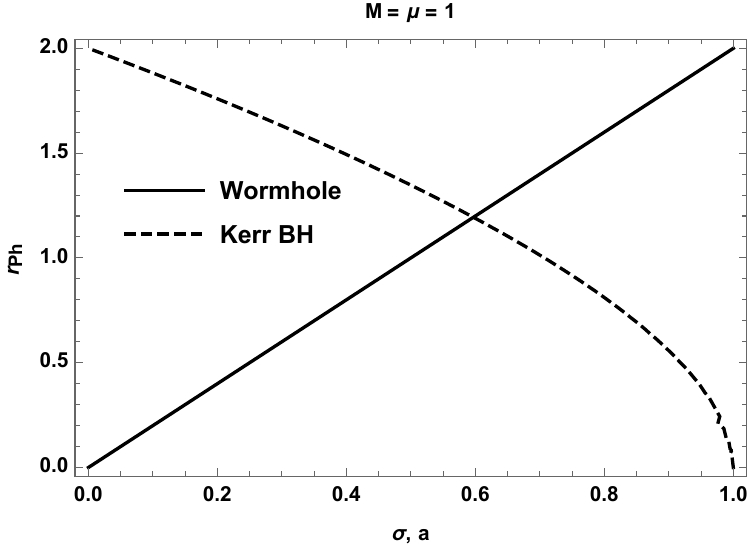}
	\centering\includegraphics[width=0.49\linewidth]{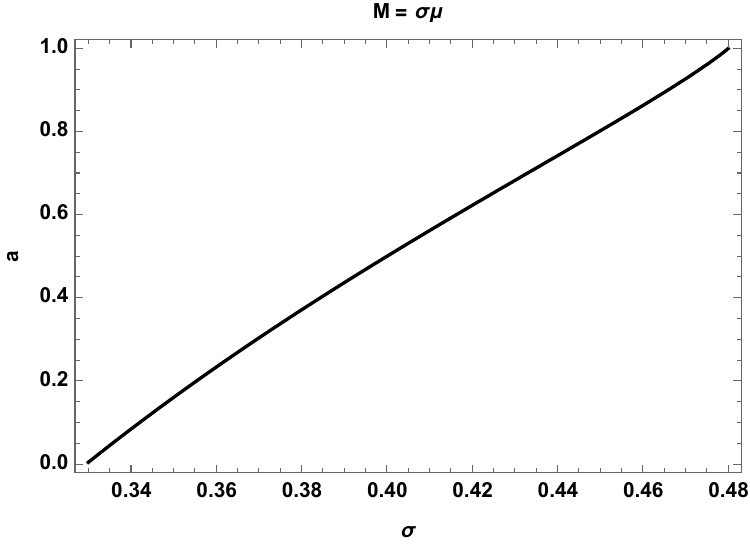}
	\caption{Dependence of the photon capture radius on the parameters $a$ and $\sigma$. The degeneracy plot between $a$ and $\sigma$ for the matching photon capture radius is shown, and a comparison is made between the photon capture radius $r_{ph}$ in the ZV wormhole space-time and the Kerr black hole. Left panel: value of $r_{ph}$ (in the rescaled coordinates) as a function of $\sigma$ for the ZV wormhole and $a$ for Kerr. Right panel: degeneracy between the parameters $\sigma$ and $a$ Kerr, i.e. the values of the two parameters that produce the same value for $R_{ph}$.
	\label{phsph}}
\end{figure*}

The dependence of the photon capture radius from the deviation parameter is illustrated in the left panel of Fig.~\ref{phsph} in comparison with the corresponding radius for the Kerr black hole as a function of $a$. 
{As in the case of the ISCO radius we can construct a definition of the photon capture radius that describes its distance from the center as 
$$
R_{ph}=\frac{l_\phi}{2 \pi}=\sqrt{g_{\phi \phi}}|_{r=r_{PH}} \; .
$$ 
Then we can obtain the degeneracy between $a$ and $\sigma$ that provides the same value of $R_{ph}$ as shown in the right panel of Fig.~\ref{phsph}.}

We again see that the deviation parameter of the wormhole metric is able to mimic the rotation parameter of the Kerr metric for the case of photon motion. 
This again leads to the conclusion that it would not be possible to distinguish a Kerr black hole from a ZV wormhole from a single measurement of the photon capture radius. However, the degeneracy of $R_{ph}$ between $\sigma$ and $a$, shown in the right panel of Fig.~\ref{phsph}, differs from that of the ISCO radius and the radiative efficiency, suggesting that a simultaneous measurement of the ISCO, radiative efficiency, and photon sphere could allow one to distinguish the two geometries.

While estimations of the radiative efficiency of accretion disks for supermassive black hole candidates can be obtained from the emission spectrum of the accretion disk, precise independent measurements of the ISCO radius and photon capture radius are not currently available. However, the imaging of the shadow of the supermassive black hole candidate at the center of the galaxy M87 suggests that such kinds of measurements, at least for one object, may become available in the near future.

The study of the motion of massless particles is also important for determining the gravitational lensing effects of the object. This is another way by which an exotic source such as the ZV wormhole may mimic the appearance of a black hole.
As mentioned, in the coordinates used for the line element in equation \eqref{metric} it is not immediately clear that the geometry becomes Minkowski at spatial infinity. This issue is easily solved by rescaling the radial and time coordinates as $r\rightarrow \xi r$ and $t\rightarrow t/\xi$, with $\xi=e^{\sigma\pi/2}$.
From the Hamilton-Jacobi equation (\ref{mHJ}) one can then write the equations of motion for the radial $r(\tau)$ and latitudinal $\phi(\tau)$ components of the geodesics in the equatorial plane $\theta=\pi/2$ as

\bea
\dot{r}^2&=& \left(\frac{\mu^2+r^2\xi^2}{r^2\xi^2}\right)^{\sigma^2+1}\left({\mathcal{E}}^2
-\frac{\mathcal{L}^2 e^{4 \sigma\tan^{-1}\left(r \xi/\mu \right)}}
{\xi^2\left(\mu ^2+r^2 \xi^2\right)} \right) , \\
\dot{\phi}&=&\frac{\mathcal{L} e^{2 \sigma \tan ^{-1}(r\xi/\mu)}}{\mu ^2+r^2 \xi^2 } .
\eea
The deflection angle $\phi$ in terms of the radial coordinate $r$ is then obtained from the integration of $d\phi/dr=\dot{\phi}/\dot{r}$.
By introducing the new variable $u=1/r$ we get the following expression for $\phi$

\bea
\frac{d\phi}{du}&=&\frac{b e^{2 \sigma  \cot ^{-1}\left(\mu u/\xi\right)}}{\left(1+\frac{\mu ^2 u^2}{\xi^2}\right)\sqrt{\left(1+\frac{\mu ^2 u^2}{\xi^2}\right)^{\sigma ^2}}} \\ \nonumber 
&&\times\frac{1}{\sqrt{ \xi^4 \left(1+\frac{\mu ^2 u^2}{\xi^2}\right)-b^2 u^2 e^{4 \sigma  \cot ^{-1}\left(\mu u/\xi\right)}}} \; .
\eea

Here $b=\mathcal{L}/\mathcal{E}$ defines the impact parameter of the photon. 
The bending angle of the photon approaching the central wormhole from infinity and going to infinity can then be found from the following integration
\begin{widetext}
	\begin{eqnarray}
\delta=2  \int_0^{1/b} \frac{b e^{2 \sigma  \cot ^{-1}\left(\mu u/\xi\right)}}{\left(1+\frac{\mu ^2 u^2}{\xi^2}\right) \sqrt{ \left(1+\frac{\mu ^2 u^2}{\xi^2}\right)^{\sigma ^2} \Big[\xi^4 \left(1+\frac{\mu ^2 u^2}{\xi^2}\right)-b^2 u^2 e^{4 \sigma  \cot ^{-1}\left(\mu
			 u/\xi\right)} \Big]}} du - \pi\ .
\end{eqnarray}
\end{widetext}

\begin{figure}[bbb]
	\centering\includegraphics[width=0.95\linewidth]{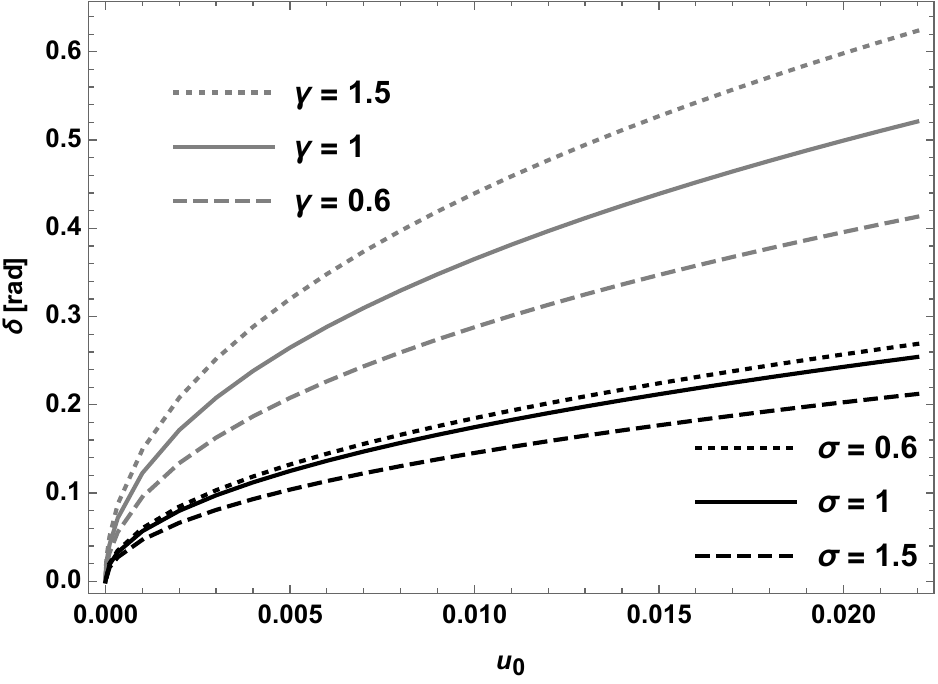}
	\caption{The dependence of the photons bending angle on the inverse of impact parameter, $u_0$ for the ZV wormhole (black lines) for various values of $\sigma$ is compared to the corresponding angle in the $\gamma$ metric for various values of $\gamma$. 
	\label{delta}}
\end{figure}

If one denotes $1/b$ as $u_0$, then the dependence of the bending angle from $u_0$ can be plotted based on the numerical calculations as in Fig.~\ref{delta}.

In Fig.~\ref{delta} we show the dependence of the bending angle $\delta$ from the inverse of the impact parameter $u_0=1/b$ for various values of $\sigma$. As $u_0\rightarrow 0$ (i.e. $b\rightarrow +\infty$) the bending angle goes to zero as expected. However $\delta$ is much smaller for the wormhole case with respect to the Schwarzschild black hole ($\gamma=1$) and $\gamma$ metric for all values of $\sigma$.
Interestingly, we also notice that the bending angle decreases for larger values of $\sigma$, while for the $\gamma$ metric the opposite is true and larger values of $\gamma$ (corresponding to oblate objects with larger quadrupole moment) produce a larger bending angle. This suggests that a simultaneous observation of deflection of light rays (i.e. a measurement of $\delta$) and the spectrum of the accretion disk (i.e. a measurement of the ISCO radius) could in principle determine whether the central object is a black hole, an exotic compact object with quadrupole moment, or a ZV wormhole.

\section{Conclusion}\label{conc}

We explored properties of a solution of Einstein's equations describing a static and axially symmetric wormhole (ZV wormhole) and investigated whether it would be possible to distinguish this source from a black hole (Kerr) or a static axially symmetric compact object with quadrupole moment ($\gamma$ metric) via the standard tools of particle motion on accretion disks and the deflection of light rays. 

The solutions considered depend on two parameters that are related to mass and quadrupole moment for the ZV wormhole and the $\gamma$ metric and mass and angular momentum for the Kerr black hole. The study of the motion of massive test particles in the equatorial plane provides the value of the innermost stable circular orbits which represents the inner edge of the accretion disk surrounding the object. The study of the motion of massless particles in the equatorial plane provides the radius of the photon capture orbit which is related to the image of the shadow of the object and the deflection angle of light-rays coming from distant sources.

{We showed that, for certain ranges of parameters, such as co-rotating accretion disks around a Kerr black hole (i.e. $a>0$) an individual measurement of the ISCO radius { or the radiative efficiency} or the photon capture radius or the deflection angle would not suffice in distinguishing between two sources. For example, from the test particle motion around the wormhole space-time it has been shown that the deviation parameter $\sigma$ can mimic the rotation parameter of the Kerr black hole.
However, simultaneous independent measurements of multiple quantities can, in principle, break the degeneracy and determine the geometry surrounding the object.}

Of course the times are not yet mature enough to practically perform such measurements for astrophysical sources, as the available data are currently too sparse and often able to provide only one of the above mentioned quantities for each candidate. However, we are reasonably hopeful that as more data are obtained
it will soon be possible to test the nature of the geometry of compact gravitational objects and determine whether they are in fact well described by the Kerr solution.

\section*{Acknowledgement}
B.N. acknowledges support from the China Scholarship Council (CSC), grant No.~2018DFH009013. AA is supported by the PIFI fund of Chinese Academy of Sciences. This research is supported in part by Grants of the Uzbekistan Ministry for Innovative Development, and by
the Abdus Salam International Centre for Theoretical Physics through Grant No.
OEA-NT-01.

\bibliographystyle{apsrev4-1}
\bibliography{gravreferences}

\end{document}